\documentclass{article}

\usepackage{arxiv}

\usepackage[utf8]{inputenc} 
\usepackage[T1]{fontenc}    
\usepackage{hyperref}       
\usepackage{url}            
\usepackage{booktabs}       
\usepackage{amsfonts}       
\usepackage{nicefrac}       
\usepackage{microtype}      
\usepackage{threeparttable}
\usepackage[numbers, sort, comma, square]{natbib}

\title{Authorship ethics: an overview of research on the state of practice}

\author{
  Nasir Mehmood Minhas \\
  SERL Sweden,
		Blekinge Institute of Technology,
		Karlskrona, Sweden \\
  \texttt{nasir.mehmood.minhas@bth.se} 
}

\begin{document}
\maketitle

\begin{abstract}
Authorship ethics is a central topic of discussion in research ethics fora. There are various guidelines for authorship (i.e., naming and order).  It is not easy to decide the authorship in the presence of varying authorship guidelines. 
This paper gives an overview of research on authorship practices and issues. It presents a review of 16 empirical research papers published between 2014 -- 2020. The objective is to learn how various research disciplines handle authorship. What are the authorship practices in various research disciplines, and what are the issues associated with these practices? 
\end{abstract}


\section{Introduction}
\label{sec:int}
The ethics of authorship is well-researched in the health sciences, but it has got less attention from social scientists \cite{macfarlane2017ethics}. In higher education and research institutions, authorship credit is considered academic currency \cite{bennett2003unethical,bulow2018hostage}. Several organizations, especially the publishers of scientific articles, provide guidelines on authorship credit. For instance, in health sciences and its related disciplines, the International Committee of Medical Journal Editors \href{http://www.icmje.org/recommendations/browse/roles-and-responsibilities/defining-the-role-of-authors-and-contributors.html}{(ICMJE)}\footnote{http://www.icmje.org} has published the authorship criteria (see Table \ref{tab:C-ICMJE}) as a part of Vancouver protocol. Vancouver protocol includes the guidelines and recommendations for conducting, reporting, editing, and publication of scholarly work in medical journals. The influence of ICMJE criteria is expanding to other disciplines as well \cite{chang2019definition}. The authorship refers to two essential aspects that the research team needs to decide about, 1) Who should be included as an author? (Authorship Naming), 2) Who would be listed as the first author, and what would be the subsequent positions of co-authors? (Authorship order). Authorship guidelines suggest that a substantial contributor would be listed as an author. However, in which order, the authors' names should be displayed, guidelines are silent in this regard \cite{smith2019misconduct,mavis2019authorship}. Other contributors who do not meet authorship criteria, their names should have to be listed in the acknowledgment section\cite{chang2019definition}.

The guidelines on who should be an author differ from one organization to another, and the difference could be within the same research discipline. These differences could be confusing for the authors and ultimately could lead to poor authorship decisions \cite{bovsnjak2012prescribed}. The existing authorship criteria provide a road-map about authorship naming, but it is hard to find such guidelines regarding authorship order \cite{helgesson2019authorship}. 
Although almost every scientific article publisher provides authorship criteria, researchers reported disagreements and unethical practices regarding authorship, for instance, giving authorship to undeserving persons or ignoring substantial contributors \cite{huth1986irresponsible}.
There is a need to increase awareness about the ethics of authorship and open up a dialog among the institutions and researchers \cite{claxton2005scientific2}. 

This study aims at investigating the practices and issues related to the ethics of authorship. A review of empirical articles/papers published between 2014 to 2020 has been conducted. The selection of the primary studies was made based on the criteria that the presented study should be discussing the authorship criteria, practices, or issues from the perspective of research ethics, and it should be a primary study. Hence, the selected primary studies represent the opinion and experience of researchers regarding the authorship criteria, practices, and challenges and issues. The selected primary studies are mainly from health science and its allied disciplines. 

The remainder of this paper is organized as follows: Section \ref{sec:rw} provides the summaries of existing reviews on the topic of authorship,  Section \ref{sec:res} presents the findings of this study regarding the authorship criteria/practices, and Section \ref{sec:con} provides the discussions and conclusions regarding research questions.

\begin{table}[!htb]
    \begin{center}
 \footnotesize
  \caption{ICMJE authorship criteria}
    \begin{tabular}{cp{7cm}}
     \hline
    ID & Criterion Description \\
     \hline
        C1& Substantial contribution to the conception or design of the work; or the acquisition, analysis, or interpretation of data for
the work  \\
        C2& Drafting the work or revising it critically for important intellectual content\\
        C3& Final approval of the version to be published\\
        C4& Agreement to be accountable for all aspects of the work in ensuring that questions related to the
accuracy or integrity of any part of the work are appropriately investigated and resolved\\  \hline
    \end{tabular}
    \begin{tablenotes}
            \centering \item To be an author one should have to fulfill: (C1 \& C2 \& C3 \& C4)
            
           \centering \item {International Committee of Medical Journal Editors, 2019}.	  
		\end{tablenotes}
    \label{tab:C-ICMJE}
     \end{center}
\end{table}

\section{Related Work}
\label{sec:rw}
From 2014 to 2021, ten review studies found to be closely related to the topic of authorship issues and practices \cite{bulow2018hostage,helgesson2019authorship,johal2017political,kornhaber2015ongoing,kumar2018ethical,mandal2015scientific,ranieri2019questionable,selbach2018academic,tang2018responding, mansourzadeh2020survey}. The selection criteria for the secondary studies was that the selected study should be a review article/paper on the topics relating to authorship criteria, authorship practices, or authorship issues.
A summary of each secondary study included in this paper is provided in the following paragraphs.

B{\"u}low and Helgesson \cite{bulow2018hostage} discussed the practices related to gift authorship, honorary authorship, and hostage authorship. These practices refer to including a person as an author who does not qualify to be an author. The authors discussed the probable motives of such practices and termed these practices as severe ethical concerns. Furthermore, they argue that accepting the authorship request of an undeserving person in a hostage-like situation resembles the problem of dirty hands (A term used in political science and political philosophy). 
The authors suggest that, if it is unavoidable and essential to carry out the research, one may accept to include the undeserving person as an author. Tang \cite{tang2018responding} criticizes opting for dirty hands even for bigger benefits for the researchers and the research community. 
Ranieri \cite{ranieri2019questionable} further discusses it from the point of view of junior researchers. Ranieri agrees with B{\"u}low and Helgesson in their opinion of getting hands dirty in a hostage-like situation. The author thinks that junior researchers could be very much in a hostage-like situation related to their supervisors, which could be an unavoidable situation to them. 

Helgesson and Eriksson \cite{helgesson2019authorship}, reviewed the practice of authorship order from the perspective of research ethics. They concluded that there is no common understanding of authorship order across disciplines, and authorship guidelines do not provide any significant criteria regarding authorship order. The authors argued that the systems based on fixed value assignment for assessing the authorship position are misleading and unfair. 
Johal et al. \cite{johal2017political}, highlighted the issues of politically motivated co-authorship. They described that some authors could be listed as co-authors, without any contribution, because of their influence in research. This would be done because of some political motives, for instance, career advancement, increasing the value of article/project by adding senior researchers, etc. The authors suggested that such practices should be avoided and considered as corrupt practices.   

Kornhaber et al. \cite{kornhaber2015ongoing}, conducted a review to explore the practices and issues regarding authorship. They found that honorary (guest/gift) authorship and ghost authorship are the key authorship issues in health sciences research, and awarding equal authorship is an increasing trend.    
Kumar \cite{kumar2018ethical} identified ghost authorship and honorary authorship as concerns to research integrity. Mandal et al. \cite{mandal2015scientific} termed the issues of honorary authorship and ghost authorship as authorship misconduct. They have provided evidence that a sizeable number of articles suffer from these problems. The authors also highlighted the conflicts regarding the authorship order and pointed out that giving equal credit to the authors is a growing trend. 
Selbach et al. \cite{selbach2018academic} presented a review of the research to investigate the issue of authorship and co-authorship. They have highlighted the unethical authorship practices, including gift, ghost, and guest authorship. The authors conclude that co-authorship credit is affected by power relations and personal interests. 
Mansourzadeh et al. \cite{mansourzadeh2020survey} surveyed Iranian retracted publications indexed in pubmed. The authors revealed that more than half of the publications are retracted because of authorship and plagiarism issues. 

The existing reviews stressed that practices like gift authorship, guest authorship, and ghost authorship affect the integrity of collaborative research. By complementing the presented related work, this study aims to understand the authorship practices in different disciplines and highlight the issues concerning research ethics.

\section{Methodology} 
\label{sec:method}
\subsection{Aim and objectives}
This review article aims to investigate the authorship issues and practices from the perspective of research ethics. To fulfill the stated aim, the following are the objectives:
\begin{enumerate}
    \item To understand the authorship criteria across the disciplines
    \item To understand the authorship practices in different research disciplines
    \item To understand the issues related to authorship naming and ordering
\end{enumerate}
\subsection{Research questions} In the light of defied objectives, the following research questions were investigated:
\begin{enumerate}
   
     \item[RQ1] What have researchers reported about the state of practice regarding authorship naming and authorship order? 
    \item[RQ2] What are the issues and unethical authorship practices reported in the literature?
\end{enumerate}

 \begin{table}[!htb]
\caption{Inclusion Criteria}
     \centering
   \footnotesize
     \begin{tabular}{cp{7cm}}
     \hline
        ID  & Criterion \\ \hline
         IC-1 & The main focus of the selected article should be on research ethics\\
         IC-2& The article must be discussing the authorship issues and practice\\
         IC-3& The selected article should represent an empirical study (Survey and Case Study) \\
         IC-4 & The selected article should be published from 2014 to 2020\\
         IC-5& The selected article should be in English\\\hline
     \end{tabular}
     
     \label{tab:IC}
 \end{table}
 
  \begin{table}[!htb]
 \caption{Exclusion Criteria}
     \centering
    \footnotesize
     \begin{tabular}{cp{7cm}}
     \hline
        ID  & Criterion \\ \hline
         EC-1 & If the main focus of the selected article is not on the research ethics\\
         EC-2& If the article is discussing issues other than authorship\\
         EC-3& If it is a review article\\
         EC-4 & If the article was published before 2014\\
         EC-5& If the article is written in the languages other than English\\
         EC-6& If the full text of the article is not available\\ \hline
     \end{tabular}
     \label{tab:EC}
 \end{table}
\subsection{Study selection}
For the study selection, I applied the search string (\textit{( "research ethics" )  AND  ( "authorship ethics"  OR  "authorship perception"  OR  "authorship issues"  OR  "authorship practices"  OR  "co-authorship"  OR  "authorship credit"  OR  "authorship criteria" ))}) in Scopus\footnote{www.scopus.com} for the year 2014 and 2020. Database searches returned a total of 180 articles, and after the title screening, 43 articles were retrieved. After reading the abstract and applying the inclusion criteria (See Tables \ref{tab:IC} and \ref{tab:EC}),  16 primary studies were selected for further review and analysis.  

\begin{table*}[!ht]
\footnotesize
	\centering
	\caption{Data extraction form}
	\begin{tabular}{lp{4cm}p{10cm}}
		\hline
		No. & Item & Data extracted \\
		\hline
		1& General information & Study title, abstract, publication year, venue, country \\
		2& Author affiliations & Departments and/or faculties of the study authors \\
		3& Study participants & Early career, mid career, or senior researchers \\
		4& Field& Field of research where study has been conducted (e.g., Medical, social science, etc.)\\
		5& Study type & Case study, experiment or survey, or other \\
		6& Focus & Primary focus of the investigation \\
		7& Findings& Key findings of the study\\ 
		8& Challenges and issues & Specific challenges and/or issues regarding authorship ethics highlighted in the findings of the study (e.g., authorship naming and ordering, definition of authorship, etc.) \\
		 \hline
	\end{tabular}
	\label{tab:extraction}
\end{table*}
 
 \subsection{Data extraction} Table \ref{tab:extraction} presents the data extraction form. The information extracted from the primary studies was stored in the excel sheets. The extracted information includes metadata of included primary studies, authors data, subjects of the study. The focus was to extract the data related to the study's primary focus, key findings, and challenges and issues concerning the ethics of authorship.  
 
 \subsection{Study limitations}
The representativeness of the findings could be a limitation of this study as the searches performed in only a single database (Scopus) and are time-bounded. Therefore the author does not claim exhaustive searches of the primary studies. However, the consistency of findings somehow minimizes the study's limitation. Another limitation of this study could be the quality of results and the researcher's bias. This limitation is minimized by involving two senior researchers who reviewed the results and provided their feedback.            

\begin{table*}[!ht]
\scriptsize
\centering
     \caption{Included primary studies}
     \begin{threeparttable}
    \begin{tabular}{|p{2.6cm}|p{5cm}|p{5cm}|p{1cm}|p{1.25cm}|p{0.5cm}|}
    \hline
        Study ID &Title& Focus&Method& Field\footnotemark[1]& Year  \\ \hline
        \citet{decullier2020have}&Have ignorance and abuse of authorship criteria decreased over the past 15 years?& To study the awareness of clinical researchers about the authorship ethics and ICMJE criteria. & Survey & Medical & 2020\\
         \citet{smith2019misconduct}& Misconduct and Misbehavior Related to Authorship Disagreements in Collaborative Science& Disagreements regarding authorship naming and ordering. Factors that cause disagreements, and misconduct and misbehavior associated with these disagreements.&Survey&Science \& Engineering&2019\\
         \citet{rees2019importance}&Importance of authorship and inappropriate authorship assignment in paediatric research in low- and middle-income countries&To understand the similarities and differences among the researchers of pediatrics science from low- and middle-income countries and high income countries regarding the importance of authorship position, and inappropriate authorship assignments.&Survey&Medical&2019\\
         \citet{breet2018academic}&Academic and Scientific Authorship Practices: A Survey Among South African Researchers&To investigate the understanding of the researchers regarding authorship criteria and gauge their ability to apply it in the multi-authored studies.&Survey&Science \& Engineering&2018\\ 
         \citet{helgesson2018misuse}&Misuse of Co-authorship in Medical Theses in Sweden&To explore the issues of authorship that have encountered by the researchers who have recently completed their doctoral thesis. &Survey&Medical&2018\\
         \citet{alshogran2018understanding}&Understanding of International Committee of Medical Journal Editors Authorship Criteria Among Faculty Members of Pharmacy and Other Health Sciences in Jordan& To evaluate the perceptions, attitudes, and practices of researchers concerning authorship criteria. &Survey&Medical&2018\\
         \citet{uijtdehaage2018whose}&Whose paper is it anyway? Authorship criteria according to established scholars in health professions education&To represent the opinion of leading researchers regarding an appropriate criteria of authorship.&Survey&Medical&2018\\
         \citet{mavis2019authorship}&Authorship Order in Medical Education Publications: In Search of Practical Guidance for the Community&To study the practices regarding authorship order and to learn the way researchers from US medical education decide the authorship order.&Survey&Medical&2018\\ 
         \citet{macfarlane2017ethics}&The ethics of multiple authorship: power, performativity and the gift economy&To study the attitude of researchers representing higher education faculties regarding multiple authorship.& Case-based Survey&Education (Social Sciences)&2017\\
         \citet{elliott2017honorary}&Honorary Authorship Practices in Environmental Science Teams: Structural and Cultural Factors and Solutions&To represent the opinion of researchers regarding authorship policies and practices.&Interview-based Survey&Environmental Sciences&2017\\
        \citet{kumar2016perceptions}&Perceptions of scholars in the field of economics on co-authorship associations: Evidence from an international survey&To know the perceptions of economics researchers regarding co-authorship, authorship ordering and authorship practices. &Survey &Economics (Social Sciences) &2016\\
        \citet{bozeman2016trouble}&Trouble in Paradise: Problems in Academic Research Co-authoring&To investigate the exclusion of the deserving individuals from co-authorship and unfair inclusion of co-authors.&Interview based study&Science \& Engineering&2016\\
        \citet{das2016knowledge}&Knowledge on ethical authorship: A comparative study between medical and pharmacy faculty&To access the understanding of Indian health sciences researchers regarding the ethics of authorship.&Survey&Medical&2016\\
        \citet{vsupak2015icmje}&ICMJE authorship criteria are not met in a substantial proportion of manuscripts submitted to Biochemia Medica& To assess the level of implementation of ICMJE criteria and to investigate the authorship violations among the researchers from medical sciences.&Survey&Medical&2015\\
        \citet{seeman2015authorship}&Authorship Issues and Conflict in the U.S. Academic Chemical Community&To investigate the authorship conflicts between early career researchers and their seniors.&Survey&Chemistry&2015\\
        \citet{al2014honorary}&Honorary authorship in biomedical journals: How common is it and why does it exist?&To determine the presence of honorary authorship in biomedical publications and identify the factors that lead to its existence.&Survey&Medical&2014\\
         \hline
    \end{tabular}
   \begin{tablenotes}
            \item [1] Medical represents all the disciplines that are related to medicine and health sciences.

		\end{tablenotes}
		\end{threeparttable}
    \label{tab:PS}
\end{table*}


\section{Results}
\label{sec:res}
Table \ref{tab:PS} presents a summary of the primary studies reviewed in this report, and Table \ref{tab:issues} presents the definitions of key terms found in the selected primary studies. A majority of the primary studies (9 of 16) are representing the medical (all disciplines related to health sciences), followed by science and engineering (3 of 16), then social sciences (2 of 16), and finally, one study each from chemistry and environmental sciences. The findings of this review are presented in subsections \ref{SubSec:Cri} and \ref{SubSec:Prac}. 

\begin{table*}[!htb]
    \centering
    \footnotesize
     \caption{Definition of terms concerning authorship practices}
    \begin{tabular}{lp{9cm}l}
   
      \hline  Issue &Definition&Ref  \\  \hline
        Authorship& Authorship refers to identifying individuals who have a substantial contribution to the conducting and reporting of research and stand responsible for the work while receiving due credit.&\cite{mavis2019authorship,smith2019misconduct,vsupak2015icmje}\\
        Ghost authorship & Denying authorship to one who has significant contributions in the study or who fulfill the authorship criteria.& \cite{uijtdehaage2018whose,helgesson2018misuse,bozeman2016trouble,seeman2015authorship,macfarlane2017ethics,alshogran2018understanding}\\  
        Gift authorship & Granting authorship as favor to one who does not qualifies the authorship criteria. &\cite{uijtdehaage2018whose,smith2019misconduct,macfarlane2017ethics,helgesson2018misuse,das2016knowledge,alshogran2018understanding}\\  
        Guest/Honorary authorship& Granting authorship to senior individuals as courtesy or to add prestige to the paper. &\cite{uijtdehaage2018whose,helgesson2018misuse,elliott2017honorary,kumar2016perceptions,bozeman2016trouble,al2014honorary,alshogran2018understanding,decullier2020have}\\  
        Hostage authorship& Including an undeserving individual as author because of unavoidable pressure. & \cite{macfarlane2017ethics,das2016knowledge,alshogran2018understanding,uijtdehaage2018whose}\\  \hline
       
    \end{tabular}
   
    \label{tab:issues}
\end{table*}

\subsection{Authorship criteria} 
\label{SubSec:Cri}
A person would be listed as an author in a scientific paper if she has a significant contribution, and it is essential to measure the contribution based on some defined criteria. Authorship criteria provide the basis to assess the contribution of an individual working in a research team. Various publishers of scientific journals provide guidelines regarding authorship criteria. Variations exist in the authorship criteria across disciplines and publishers. Such differences are even visible for publishers within the same subject \cite{uijtdehaage2018whose,smith2019misconduct}. A
 majority of the authors of our selected studies referred to the authorship criteria published by ICMJE (see Table \ref{tab:C-ICMJE})  \cite{smith2019misconduct,rees2019importance,breet2018academic,helgesson2018misuse,alshogran2018understanding,uijtdehaage2018whose,macfarlane2017ethics,das2016knowledge,vsupak2015icmje}. The ICMJE criteria are accepted worldwide by the journals from medicine and related health sciences disciplines. The authors from the disciplines other than medicine and health sciences are also referring to the ICMJE criteria \cite{smith2019misconduct, breet2018academic,macfarlane2017ethics}. 

Authorship naming and order are the two primary aspects, and the research team needs to decide carefully about authorship naming and order. Poor or careless decision-making can lead to disagreements and unethical practices. The following subsections discuss the state of practice regarding authorship naming and order.  

\subsubsection{Authorship naming criteria: State of practice}
Many authors reported that authorship criteria are not followed in their complete letter and spirit while naming an author. Some of the instances are reported in the following paragraphs. 

Helgesson et al. \cite{helgesson2018misuse} surveyed with the researchers who have recently completed their doctoral degrees in medicine. The authors investigated the authorship issues concerning compliance to the authorship guidelines. According to the majority (53\%) of the respondents, ICMJE authorship guidelines were not respected in at least one of their articles. However, a vast majority (80\%) of the respondents reported their awareness of authorship guidelines. Breet et al. \cite{breet2018academic} conducted a survey with the science and engineering researchers of South Africa and found that 80\% of the survey respondents are aware of authorship guidelines 52\% confirmed that they found it easy to implement the authorship criteria. 
Alshogran et al. \cite{alshogran2018understanding} investigated the understanding of ICMJE authorship criteria among the health sciences researchers of Jordan. The authors revealed that only 27\% of the respondents reported their awareness about the ICMJE guidelines. However, most of the respondents (77\%) agreed that authorship should be granted only if one qualifies the ICMJE authorship criteria. In a study with the research leaders from the health profession, Uijtdehaage et al. \cite{uijtdehaage2018whose} reported that 58\% of the respondents correctly identified the ICMJE authorship criteria. There were some assumptions of the senior researchers regarding authorship. For instance, 14\% of senior researchers considered supervising or leading a research team qualifies one to be an author. Similarly, 14\% of them believed that drafting a manuscript should not be regarded as a criterion of authorship. According to Uijtdehaage et al.,  these assumptions are not aligned with the defined standards. Macfarlane \cite{macfarlane2017ethics} conducted a case-based survey to understand the viewpoint of academics from social sciences regarding ICMJE authorship criteria. The results revealed that 77\% of the participants agreed with the first criterion of ICMJE, 48\% were in favor of the second criterion, 77\% supported the third criterion, and only 13\% thought that the fourth criterion is sufficient to be an author. The ultimate perception of the participants was, anyone who has a substantial contribution is a legitimate author, but the most difficult part is to define the substantial contribution. In an interview-based study with six research teams from environmental sciences, Elliott et al. \cite{elliott2017honorary} revealed that half of the participants are using formal written policies and the other half are using informal procedures to decide who among the team members should be named as an author.  The authors reported that these procedures (formal or informal) are overinclusive. The primary cause of being overinclusive is the difficulties in determining the extent of the contribution. In a survey with the medical and pharmacy faculties of India \cite{das2016knowledge}, it is reported that only 22\% of medical and 39\% of pharmacy faculty members were aware of COPE (Committee On Publication Ethics) or ICMJE guidelines on authorship. In a survey of articles submitted with a medical journal (2013 to 2015), \citet{vsupak2015icmje} analyzed the self-reported contribution statements to assess the conformance with ICMJE criteria. They reported that 61\% of the researchers were utterly following the ICMJE criteria, while 9\% of them were not following it at all.  Furthermore, it is revealed that in the self-reported contribution statements where the authors claimed to fulfill the authorship criteria, only 49\% of them were qualified according to ICMJE criteria. 

From the discussion on the findings of existing primary studies, it could be concluded that the authorship criteria are not followed in the majority of the cases, and there is a lack of awareness regarding the authorship criteria. 
\subsubsection{Authorship order: State of practice} 
Authorship order is translated differently in different academic / research disciplines. For instance, in some research areas, the authors' names are listed in alphabetical order. In many research domains, the authorship order exhibits the authors' contribution to the study. Various research and academic institutions use the authorship order to determine the scientific merit of the individuals \cite{smith2019misconduct}.  
 
In a survey-based study, Rees et al. \cite{rees2019importance} revealed that there is a contrast between the authors from low-and-middle-income countries (LMIC) and high-income countries (HIC) regarding authorship order. Authors representing LMIC perceive the first position as the most important, whereas HIC representatives consider the last position as most significant.  In a survey with the Jordanian health sciences, researchers Alshogran et al. \cite{alshogran2018understanding} revealed that 79\% of respondents agreed that the authorship order should be based on the respective contributions, and the decision should be taken jointly by all authors.  In a survey with the researchers, Mavis et al. \cite{mavis2019authorship} found that a majority of the respondents (86\%) supported that the authors should be listed in the decreasing of contribution, starting with the most substantial contributor as the first author. 95\% of the respondents emphasized an agreement between the contributors regarding the authorship order.
Furthermore, 94\% of the respondents were against the alphabetical listing of authors, and 74\% believed that first, second, and corresponding authors are the significant contributors. In a case-based survey by Macfarlane \cite{macfarlane2017ethics}, 32\% of the survey respondents suggest to award the first authorship to a person who needs it for career advancement (e.g., research student), 17\% suggested to use alphabetical order, and 6\% respondents voted to award the first authorship to the senior team member. Macfarlane coined the terms of gift ordering and power order. The first refers to decide the authors' order by career considerations instead of intellectual contributions. In contrast, the second term refers to determining the authors' order based on the hierarchy within the research team. In a multi-nation survey with the economics researchers, Kumar and Ratnavelu \cite{kumar2016perceptions} revealed that although it is believed that economics researchers follow the alphabetical order of authorship, however, to decide the authorship order, 35\% of the survey respondents measure the authors' contribution.     

The above discussion concludes that authorship order has significant importance in a majority of research/academic disciplines. The first, second, and corresponding authors are considered to be leading contributors in an article. In most cases, the authorship order is decided based on the respective contribution of the authors. In some disciplines (e.g., economics), authors' names are listed in alphabetical order.  

\subsection{Authorship issues}
\label{SubSec:Prac}
\subsubsection{Authorship disagreements}
In an online survey, Smith et al. \cite{smith2019misconduct} investigated authorship naming and ordering disagreements. 46\% of the respondents reported the disagreements regarding authorship naming, whereas 38\% of them described the disagreements concerning authorship order. Furthermore, it is revealed that the likelihood of disagreements varies across the disciplines, career levels, and gender. For instance, mid-career researchers are more involved in disagreements than senior researchers, and early career researchers are rarely involved in such issues. On the gender side, compared to male researchers, most of the events of disagreements were reported by the female researchers. Results show that for authorship naming, the most critical factor that causes disagreements is valuing an individual's contribution. The cause of these disagreements could be differing ethics, lack of discussions and agreement within the team, and lack of clarity about criteria. 
Breet et al. \cite{breet2018academic} highlighted the researchers' experience regarding the authorship disagreements, 59\% of their survey respondents confirmed that they had encountered disagreements regarding authorship naming, and 48\% of the respondents reported such disagreements in authorship order. The survey revealed that most disagreements were due to different ways of measuring the contributions. Furthermore, female researchers were highlighting more disagreements concerning authorship order compared to authorship naming.  Alshogran et al. \cite{alshogran2018understanding} reported that a majority of the respondents of their survey-based study reported disagreements on authorship. Based on the experiences of the researchers, Bozeman and Jan \cite{bozeman2016trouble} discussed the events of disagreements in the authorship order. In the reported incidents, the junior or middle career researchers were the complainants against the senior researchers. Seeman and House \cite{seeman2015authorship} reported authorship conflicts among the junior researchers and their seniors, colleagues from the same institution, and colleagues from different institutions. A small number of affectees took their concerns to the offending individuals, while most did not take this initiative because of fear of not getting a positive outcome and being in an undermined situation.  

From the findings of the reviewed studies, it could be concluded that the majority of disagreements relate to authorship naming. However, such conflicts also exist regarding the authorship order. The complaining individuals are primarily early- and mid-career researchers and they have such complaints against the senior members of their research teams. Furthermore, the majority of these disagreements were reported by female researchers.
\subsubsection{Unethical authorship practices}
In a survey study \cite{rees2019importance}, it is revealed that a majority of the researchers have an opinion that, in most cases, authorship order is assigned inappropriately, and a sizeable number of researchers told that they did accept the wrong author-order to avoid the conflicts with the collaborators. Helgesson et al. \cite{helgesson2018misuse} revealed that 28\% of the survey respondents experienced the handling of authorship order without merit. Furthermore, the respondents reported the inclusion of one or more authors without having a substantial contribution. Alshogran et al. \cite{alshogran2018understanding} revealed that in their survey-based study, 30\% of the respondents claimed that they were forced to add an undeserving author. It is also reported that the incorrect placing of authors and issues of the ghost, gift, or guest authors are common in health sciences researchers of Jordan. Uijtdehaage et al. \cite{uijtdehaage2018whose} reported that 63\% of their survey respondents told that without contributing, senior-level administrators are listed as co-authors in the papers of faculty working under them. 60\% of respondents said that they witness awarding gift authorship to the individual who did not have any input in the research work. Honorary authorship was also reported by the majority (69\%) of the survey respondents, whereas, 4\% of the respondents consider the award of honorary authorship as ethical.  A large number of respondents (51\%) reported facing pressures regarding the authorship order.
Furthermore, the authors also discussed the problem of ghost authorship, not giving authorship to the contributors who fulfill the authorship criteria.  Macfarlane \cite{macfarlane2017ethics} reported the experiences of respondents regarding unethical practices in a case-based survey. The reported events were, denying authorship to a substantial contributor or given minor credit (ghost authorship), awarding authorship to supervisors who did not contribute to the work (gift authorship), and pressures from seniors/heads to include them as authors with no contribution (hostage authorship). In an interview-based study with the senior and early career researchers, Elliott et al. \cite{elliott2017honorary} reported the inappropriate authorship practices within the research teams. For instance, awarding authorship to all team members, regardless of having significant, marginal, or no contributions in the article, the authors termed the phenomenon as honorary authorship. In a multi-nation survey with the economics researchers, Kumar and Ratnavelu \cite{kumar2016perceptions} also pointed out the practice of awarding honorary authorship. In an interview-based study with science and engineering researchers, Bozeman and Jan \cite{bozeman2016trouble} shared the stories of different researchers regarding the issues of excluding the deserving individuals from the authors' list (ghost authorship) and awarding authorship to non-deserving individuals (guest authorship). In a comparative survey between medical and pharmacy faculty members in India, Das et al. \cite{das2016knowledge} revealed that 89\% of medical and 37\% of pharmacy faculty members were involved in awarding gift authorship to their professors and head of departments.
Similarly, 81\% of medical and 29\% of pharmacy faculty members were pressurized by the seniors to include their names in the articles (Hostage authorship). In a survey with the chemistry researchers from the US, Seeman and House \cite{seeman2015authorship} reported ghost authorship practice. The authors found that 50\% of the respondents were not included as a co-author, although they had contributed significantly to work. 35\% reported that their supervisor did not acknowledge their contribution. 41\% were complaining about a colleague from their institution, while 42\% said that a colleague from another institution did this. In a survey with the medical researchers, \citet{al2014honorary} investigated honorary authors and reasons to award the honorary authorship. The results revealed that 33\% of respondents awarded honorary authorship to others. The reasons for the award of honorary authorship in terms of percentage of respondents were, complimentary award by 39\%, avoiding conflicts at workplace 16\%, and to increase the chance of article acceptance 7\% 

The above discussion leads to the fact that unethical authorship practices are common in almost all research fields. Regardless of gender or experience, the researchers are reportedly involved in such activities. In some cases, the researchers were in favor of some of the practices. For instance, awarding authorship to an undeserving individual in a hostage like situation, or gifting authorship to an individual for the career benefits. 

\section{Discussion and Conclusion}
\label{sec:con}
 This paper reviewed the empirical studies on the topic of authorship published from 2014 to 2019. Given that searches are limited and may not always find all relevant studies, it is a limitation of this study. This limitation did not affect the findings since the results obtained from the papers are consistent and reveal a pattern. The selected studies investigated the state of practice regarding the ethics of authorship. The following paragraphs summarize the findings. 

Regarding RQ1 (\textit{``What have researchers reported about the state of practice regarding authorship naming and authorship order?"}), the intention was to learn about the understanding of researchers regarding authorship criteria and to investigate the current practices associated with authorship naming and authorship order. From the studies of different disciplines, it is revealed that the most followed authorship criteria is the one defined by ICMJE, and along with the medical sciences, researchers from other subjects are also referring to it.
However, studies revealed that a majority of researchers are not following the authorship criteria in their papers. One of the reasons in this regard is the awareness of the researchers, especially from low- and middle-income countries. 
Authorship criteria provide the guide to decide who should be named as an author in a scientific study. However, the criteria do not provide any support regarding the authorship order. 
 The practices related to authorship order vary among the different disciplines. For instance, in some research disciplines, authors' names are listed in alphabetical order, whereas others use the assessed value of the contribution to decide the author's order. In most cases, first, second, and corresponding author positions are considered significant. In practice, two extremes were revealed,  1) strictly following the authorship criteria, and 2) being overinclusive. In the first case, by strictly applying the ICMJE authorship criteria, there is high likelihood that some members may not be included as authors of a scientific study because of failing to meet some points of criteria. Hence, maybe denying authorship to a deserving individual. Whereas in the second case, it is revealed that every member of the research team will be included as an author at the first stage. Later, members can decide to be on the authors' list or request to pull their names from the list. Ultimately, maybe awarding authorship to an undeserving person. These two scenarios are the basis of the majority of the unethical authorship practices.  

Concerning RQ2 (\textit{``What are the issues and unethical authorship practices reported in the literature?"}), the focus was to investigate the issues regarding authorship that researchers are facing in multi-authored studies.  This report highlights two aspects, 1) authorship disagreements and 2) unethical authorship practices. The research reports various disagreements concerning authorship, mainly the conflicts relate to authorship naming and authorship order. In most cases, the conflicts were reported by early career and mid-career researchers, and they were complaining about the events of disagreements with their seniors. However, there are only a few instances where the mid-career researchers took up the conflicting issues, while most early-career researchers avoided doing so.
The stated reasons for not bringing up the disagreements were fear of affecting the working relationships with the colleagues, expectations of positive outcomes, and being in a hostile situation. It has revealed that various unethical practices are associated with authorship, for instance, awarding authorship to an undeserving individual (gift authorship, guest or honorary authorship), awarding authorship under pressure (hostage authorship), and denying authorship to a substantial contributor (ghost authorship). 
\subsection{Authorship practice in software engineering}
As a software engineering researcher, the author was interested in reviewing the studies presenting the state of authorship practice in software engineering or computer science disciplines.  The author could not find even a single study investigating the authorship practice in these disciplines. Therefore, the author could not reflect on the state of authorship practice in software engineering and computer science. However, considering the reports from other disciplines, it is assumed that there may be ethical issues associated with authorship practice in these research areas. 
Empirical investigations from these research disciplines are needed to present the state of practice on the ethics of authorship, and such studies could help create an awareness among the research community. 
\subsection{Concluding note}
From the findings of existing empirical studies on authorship, it is concluded that there is a need to create awareness among the researchers regarding authorship ethics and criteria. In this regard, universities, research institutions, and senior researchers have to take the lead. The authorship guidelines need to be published at the institutional level, and every early career researcher (e.g., Ph.D. student) should have to be provided with training on research ethics. The institutions, departments, and research groups need to define clear policy guidelines regarding authorship (naming and ordering). Considering the guidelines, at the beginning of every research venture, the research team members have to make sure to define the team's policy on authorship. And to identify the substantial contributor(s), the contribution analysis should have to be done after every phase of research work. In this regard, web-based task assignment and tracking applications can help assign and monitor the tasks to the team members.  At the departmental level, there could be an oversight committee that could help in the resolution of authorship conflicts. The editors of scientific journals may encourage the submission of a contribution statement duly signed by all contributors. 
\subsection{Future work}
The author is currently working on a survey-based study that would be conducted with software engineering researchers. The aim is to investigate authorship practices and issues.

\section*{Data Availability}
It is a review paper, and all included primary studies are listed in Table \ref{tab:PS} and are available online. Furthermore, data extracted from the included papers are comprehensively presented in Section \ref{sec:res} and discussed in Section \ref{sec:con}.
\section*{Acknowledgement} The author would like to thank Prof. J\"{u}rgen B\"{o}rstler and Prof. Claes Wohlin, for their valuable feedback on initial phases of the review and subsequently on the drafts of this paper.

\bibliographystyle{unsrtnat}
\bibliography{Ethics}

\end{document}